\documentclass[apj,chicago,twocolumn,numberedappendix,twocolappendix]{emulateapj}
\usepackage{graphicx}
\usepackage{mathrsfs}
\usepackage[intlimits,centertags]{amsmath}
\usepackage{amssymb,amsfonts}
\usepackage{enumerate}
\usepackage[pdftex]{hyperref}
\usepackage[x11names]{xcolor}
\usepackage{booktabs}
\usepackage{mathpazo}

\hypersetup{pdftitle={Searching for All-Scale Anisotropies in the Arrival Directions of Cosmic Rays above the Ankle},
pdfsubject={Searching for All-Scale Anisotropies in the Arrival Directions of Cosmic Rays above the Ankle},
pdfauthor={Markus Ahlers},
pdfstartview={FitH},
colorlinks=true,
bookmarksopen=false,
bookmarksnumbered=false,
bookmarksopenlevel=0,
linkcolor=Blue1!60!black,
citecolor=Green1!50!black,
urlcolor=Blue1!70!black
}

\begin{document}

\title{Searching for All-Scale Anisotropies\\ in the Arrival Directions of Cosmic Rays above the Ankle}
\shorttitle{Searching for All-Scale Anisotropies in the Arrival Directions of Cosmic Rays above the Ankle}

\author{Markus Ahlers}
\shortauthors{Markus Ahlers}

\email{markus.ahlers@nbi.ku.dk}
\affiliation{Niels Bohr International Academy \& Discovery Centre, Niels Bohr Institute,\\ University of Copenhagen, Blegdamsvej 17, DK-2100 Copenhagen, Denmark}

\begin{abstract}
The Pierre Auger Observatory has recently reported the detection of a dipole anisotropy in the arrival directions of cosmic rays above 8~EeV with a post-trial significance of more than $5.2\sigma$. This observation has profound consequences for the distribution and composition of candidate sources of cosmic rays above the ankle ($3-5$~EeV). In this paper we search for the presence of anisotropies on all angular scales in public Auger data. The analysis follows a likelihood-based reconstruction method that automatically accounts for variations in the observatory's angular acceptance and background rate. Our best-fit dipole anisotropy in the equatorial plane has an amplitude of $5.3\pm1.3$ percent and right ascension angle of $103\pm15$ degrees, consistent with the results of the Pierre Auger Collaboration. We do not find evidence for the presence of medium- or small-scale anisotropies. The method outlined in this paper is well-suited for the future analyses of cosmic ray anisotropies below the ankle, where cosmic ray detection in surface arrays is not fully efficient and dominated by systematic uncertainties. 
\end{abstract}

\keywords{cosmic rays --- methods: data analysis}

\maketitle

\section{Introduction}\label{sec:Introduction}

Extragalactic cosmic rays are deflected by magnetic fields during the epoch it takes to propagate from their sources to the observer~\citep{Beck:2000dc}. The spatial variation of these magnetic fields in terms of strength and orientation leads to a random walk of charged particles. This has important consequences for the study of cosmic ray sources: the arrival directions of cosmic rays are scrambled compared to those expected from rectilinear propagation and the peak arrival time is expected to be much delayed compared to the light travel time from the sources. These two effects limit the possibility to identify cosmic ray sources by cross-correlations with simultaneous emission in photons, neutrinos, or gravitational waves. In addition, the dispersion of cosmic ray arrival times can be expected to be much longer than the short emission period of transient candidate sources or the lifetime of cosmic ray observatories. These propagation effects, together with a uniform distribution of candidate sources over large distances, result in a continuous flux of cosmic rays and in arrival directions that follow a nearly isotropic distribution.

However, the non-uniform distribution of sources in the local universe can be visible by weak anisotropies in the cosmic ray arrival direction~\citep{Giler:1980dr,Harari:2013pea}. The Pierre Auger Observatory~\citep{ThePierreAuger:2015rma} has recently analyzed the arrival directions of ultra-high energy cosmic rays observed over a period of twelve years~\citep{Aab:2017tyv}. The analysis focused on data in two energy bins, 4-8~EeV and above 8~EeV, where cosmic ray detection with the Pierre Auger surface array becomes fully efficient~\citep{Aab:2014gua}. Whereas no significant anisotropy could be identified in the first energy bin, the Pierre Auger Collaboration was able to detect a dipole anisotropy above 8~EeV with a post-trial significance of $5.2\sigma$. The dipole vector has a best-fit amplitude of $(6.5^{+1.3}_{-0.9})$\% and points toward an right ascension angle $100\pm10$ degrees and declination angle $-24^{+12}_{-13}$ degrees. This observation is an important step toward the identification of ultra-high energy cosmic ray sources, and has implications for the strength of intergalactic magnetic fields, the local source density, and the chemical composition of sources~\citep{Lemoine:2009pw,Liu:2013ppa,Globus:2017fym,Wittkowski:2017nfd}.

An important limitation of the analysis method used in \cite{Aab:2017tyv} is its reliance on an accurate modeling of the detector efficiency in time and arrival direction. The systematic uncertainty of the detector exposure has been analyed by the Pierre Auger Collaboration and is expected to be below the one-percent level above 4~EeV. This is sufficiently low compared to the observed size of the large-scale dipole anisotropy observed above 8~EeV. In this paper, we will apply an alternative anisotropy reconstruction method~\citep{Ahlers:2016njl}, that is independent of an {\it a priori} detector modeling. The motivation is twofold. Firstly, the original dipole analysis by \cite{Aab:2017tyv} does not discuss the presence of medium-scale anisotropies in the cosmic ray arrival direction. Recent analyses of TeV-PeV cosmic ray data have shown that there are significant medium- and small-scale structures in the anisotropy maps~\citep{Ahlers:2016rox}. The analysis of these features after subtraction of the large-scale dipole anisotropy seems only feasible with reconstruction methods that are capable of simultaneously calibrating the detector exposure by data~\citep{Amenomori:2005pn,Amenomori:2010yr,Amenomori:2012uda,Ahlers:2016njl}. Secondly, the method discussed in this article is also well-suited for the analysis of large-scale anisotropies in Auger data below 4~EeV, where cosmic ray detection in the surface array is not fully efficient~\citep{Aab:2014gua}.

This article is organized as follows. We start in section~\ref{sec:Propagation} with a brief discussion on cosmic ray propagation in magnetic fields and the expected level of anisotropy for extragalactic sources. In section~\ref{sec:Observation} we discuss the observation of cosmic rays with ground-based observatories and introduce  the conventions and coordinate systems used for the likelihood-based reconstruction methods described in section~\ref{sec:MaxLH}. We then apply this method to publicly available Auger data in section~\ref{sec:Auger} to re-analyze the dipole anisotropy and discuss the presence of medium- and small-scale anisotropies in the residual data. Finally, we conclude in section~\ref{sec:Conclusion}. 

Throughout the paper we use Heaviside-Lorentz units and make frequent use of the abbreviation $A_x = A/(10^x u)$, where $u$ is the (canonical) unit of the quantity $A$.

\section{Cosmic Ray Propagation}\label{sec:Propagation}

A cosmic ray nucleus with charge $Z$ and momentum $p$ is deflected by magnetic fields as it propagates between the source and the observer. The maximal gyroradius of the trajectory can be expressed as $r_g \simeq \mathcal{R}/B$, where $\mathcal{R} \equiv p c/(Z e)$ is the cosmic ray rigidity and $B$ is the magnetic field strength. For cosmic rays in the ankle region ($E_{\rm CR}\simeq 3-5$~EeV) the maximal gyroradius can be estimated as
\begin{equation}
r_g \simeq 1.1{\mathcal{R}_{18}}{B^{-1}_{-6}}\,{\rm kpc}\,,
\end{equation}
where we use the abbreviation $\mathcal{R}_{18} = \mathcal{R}/(10^{18}{\rm V})$ and $B_{-6} = B/(10^{-6}{\rm G})$. These reference values correspond to the inferred magnetic field strength in the Milky Way and assume light cosmic ray nuclei ($Z\simeq1$) at the ankle.

In the presence of turbulent magnetic fields, low-rigidity cosmic ray nuclei from distant sources can be repeatedly deflected into random directions and their transport can be described as a diffusive process~\citep{1966ApJ...146..480J,1966PhFl....9.2377K,1967PhFl...10.2620H,1970ApJ...162.1049H}. The effect of random scattering in turbulent magnetic fields is encapsulated in the diffusion tensor ${\bf K}$. Standard diffusion theory predicts that the arrival directions ${\bf n}$ of cosmic rays are nearly isotropic and only perturbed by a weak dipole anisotropy $\propto \boldsymbol{\delta}\!\cdot\!{\bf n}$, which follows the gradient of the cosmic ray density, $\boldsymbol{\delta} = (3/c){\bf K}\!\cdot\!\nabla\ln n_{\rm CR}$. The peak arrival time of cosmic rays from a source at location ${\bf r}$ emitting for a short period can be estimated as $t_{\rm peak} \simeq {\bf r}^T{\bf K}^{-1}{\bf r}/6$. The time dispersion of cosmic ray arrival is expected to be of the same order, $\sigma_t \simeq t_{\rm peak}$. The diffusive regime is expected to hold for sources with a distance $d$ that is larger than the effective diffusion distance, $d \gg \lambda_{\rm diff}={\rm tr}\,{\bf K}/c$. The diffusive time dispersion is therefore $\sigma_t\gg d/c$ and we can expect large time dispersions in comparison to observational time scales. In summary, the diffusive cosmic ray regime is characterized by an isotropic distribution of cosmic ray arrival directions with a weak large-scale anisotropy and constant flux.

In the case of high-rigidity cosmic rays and close-by sources, the particle transport is not necessarily diffusive, $d\ll\lambda_{\rm diff}$. If the gyroradius is large compared to the size of the outer scale of turbulence, $\lambda\ll r_g$, we can approximate the diffusion length as $\lambda_{\rm diff} = r_g^2/\lambda$. If the source distance is still larger than the outer scale, $d\gg\lambda$, the random walk of extragalactic cosmic rays through the magnetic field with changing orientation over the length scale $\lambda$ will result in an angular and time dispersion of the signal. For random field orientations the average angular deflection over the distance $\lambda$ can be estimated as $\Delta\psi_\lambda = \sqrt{2/3}\lambda/r_g$. After propagation over $n_\lambda = d/\lambda$ cells the cumulative angular dispersion is $\Delta\psi_{\rm G} \simeq \sqrt{n_\lambda}\Delta\psi_\lambda$.

Galactic magnetic fields that extend over the Galactic halo with a half-width $H\simeq10^3$pc can be expected to be ordered over length scales $\lambda\simeq10^4$pc~\citep{Beck:2000dc}. The random walk of extragalactic cosmic rays after they entered the halo will result in an angular dispersion of the order of
\begin{equation}
\Delta \psi_{\rm G} \simeq 4.3^\circ\lambda_2^{1/2}H_3^{1/2}B_{-6}\mathcal{R}^{-1}_{18}\,,
\end{equation}
where we again use the abbreviation $\lambda_2 = \lambda/(10^2{\rm pc})$, {\it etc.} An additional contribution to the angular dispersion is expected from random deflections in intergalactic magnetic fields~\citep{Beck:2000dc}. For typical benchmark values we can estimate an angular dispersion at the level of
\begin{equation}
\Delta \psi_{\rm IG} \simeq 1.3^\circ\lambda_6^{1/2}d_7^{1/2}B_{-11}\mathcal{R}^{-1}_{18}\,,
\end{equation}
where $d$ is the distance to the source. An additional contribution to the angular dispersion can be expected from enhanced scattering in large-scale structures, {\it e.g.}, in the $10^{-7}-10^{-6}{\rm G}$ magnetic fields observed in the intracluster medium of galaxy clusters or superclusters~\citep{Xu:2005rb,Kronberg:2007wa}. The combined angular dispersion of cosmic ray events obscures the presence of close-by cosmic ray sources. Therefore, the cumulative distribution of cosmic ray arrival directions from all extragalactic sources can be expected to be nearly isotropic, but can allow for anisotropies on large and small angular scales.

The random walk through magnetic cells will also lead to a delayed cosmic ray arrival compared to the light travel time. This time delay can be estimated from the angular dispersion as $\Delta t \simeq (d/2c)(\Delta\psi)^2$~\citep{Waxman:1996zc}. The time dispersion is expected to be of similar order, $\sigma_t\simeq\Delta t$. This dispersion can only become comparable to the experimental lifetime for rare cosmic ray events with very high rigidity.  

\section{Cosmic Ray Observation}\label{sec:Observation}

For the analysis of cosmic ray anisotropies we assume that the flux of cosmic rays above the ankle can be treated as constant over the lifetime of the observatory. The angular distribution can be expressed as a function of celestial longitude $\alpha$ (right ascension) and latitude $\delta$ (declination), 
\begin{equation}\label{eq:phi}
  \phi(\alpha,\delta) = \phi^{\rm iso}I(\alpha,\delta)\,,
\end{equation} 
where $\phi^{\rm iso}$ (units of ${\rm cm}^{-2}\, {\rm s}^{-1}\, {\rm sr}^{-1}$) corresponds to the isotropic flux level, {\it i.e.}, the flux averaged over the full celestial sphere, and $I(\alpha,\delta)$ is the relative intensity of the flux as a function of position in the sky. The anisotropy is defined as the deviation $\delta I = I-1\ll 1$. 

In the local coordinate system of the ground-based observatory the arrival direction of a cosmic ray is determined by its azimuth angle $\varphi$ (from the north, increasing to the east), zenith angle $\theta$, and local sidereal time $t$. The local sidereal time is the hour angle of the zenith, {\it i.e.}, the right ascension angle of the local meridian at the time of observation. At any given time, the observatory covers an instantaneous field of view which is typically characterized by a maximal zenith angle, $\theta\leq\theta_{\rm max}$. Over every sidereal day (which is about 4 minutes shorter than the average solar day) the observatory covers an integrated field of view. For a continuously operating ground detector located at geographic latitude $\Phi$ this integrated field of view is characterized by a declination band, $\delta_{\rm min}<\delta<\delta_{\rm max}$, with $\delta_{\rm min} = {\rm max}(-90^\circ,\Phi-\theta_{\rm max})$ and $\delta_{\rm max} = {\rm min}(90^\circ,\Phi+\theta_{\rm max})$

In the following, we will assume that the detector exposure $\mathcal{E}$ per solid angle and sidereal time $t$  accumulated over many sidereal days can be expressed as a product of its angular-integrated exposure $E$ per sidereal time (units of ${\rm cm}^2\, {\rm sr}$) and relative acceptance $\mathcal{A}$ (units of ${\rm sr}^{-1}$ and normalized as $\int{\rm d}\Omega \mathcal{A}(\Omega)=1$):
\begin{equation}\label{eq:E}
  \mathcal{E}(t,\varphi,\theta) \simeq E(t)\mathcal{A}(\varphi,\theta)\,.
\end{equation}
This ansatz assumes that the relative acceptance of the detector does not strongly depend on sidereal time. This assumption is also implicit in cosmic ray background estimation by direct integration~\citep{Atkins:2003ep} or time-scrambling~\citep{Alexandreas1993}. Note that this ansatz does {\it not} imply that the detector has a constant angular acceptance over the course of many sidereal days. 

The observation in the local horizontal coordinate system is related to the cosmic ray flux in the celestial (or equatorial) coordinate system via a time-dependent transformation. We can define the unit vector ${\bf n}$ corresponding to the coordinates $(\alpha,\delta)$ in the right-handed equatorial system as
\begin{equation}
{\bf n}=(\cos\alpha\cos\delta,\sin\alpha\cos\delta,\sin{\delta})\,.
\end{equation}
Similarly, the unit vector ${\bf n}'$ corresponding to the coordinates $(\theta,\varphi)$ in the right-handed local system is 
\begin{equation}
{\bf n}'=(\cos\varphi\sin\theta,-\sin\varphi\sin\theta,\cos{\theta})\,.
\end{equation}
The two unit vectors are related via a time-dependent coordinate transformation ${\bf n}={\bf R}(t){\bf n}'$. For an experiment located at a geographic latitude $\Phi$ and longitude $\Lambda$ (measured east from Greenwich), the transformation is
\begin{equation}\label{eq:Rmatrix}
{\bf R}(t) =
\begin{pmatrix}
  -\cos \omega t\sin \Phi&-\sin \omega t\sin\Phi&\cos\Phi \\
  \sin \omega t&-\cos \omega t&0\\
  \cos \omega t\cos\Phi&\sin \omega t\cos\Phi&\sin\Phi
\end{pmatrix}\,,
\end{equation}
where $\omega = \omega_{\rm solar} + \omega_{\rm orbit}$ with solar frequency $\omega_{\rm sol} = 2\pi/24$h and Earth's orbital frequency $\omega_{\rm orbit} = 2\pi/1$yr. The local sidereal time $t$ is related to the sidereal time at Greenwich $t'$ by $t=t'+\Lambda/\omega$.

The expected number of cosmic rays at a sidereal time $t$ from an azimuth angle $\varphi$ and zenith angle $\theta$ can now be expressed as
\begin{multline}\label{eq:mu}
\mu(t,\varphi,\theta) = I(\alpha(t,\varphi,\theta),\delta(t,\varphi,\theta))\\
\cdot  \left[\Delta t \mathcal{N}(t)\right]\left[ \Delta\Omega\mathcal{A}(\varphi,\theta)\right]\,,
\end{multline}
where $\mathcal{N}(t) \equiv \phi^{\rm iso}{E}(t)$ gives the expected rate of isotropic background events at sidereal time $t$. For a known local detector acceptance $\mathcal{A}(\varphi,\theta)$ and background level $\mathcal{N}(t)$, the previous relation allows us to reconstruct the relative cosmic ray intensity $I$ and the cosmic ray anisotropy $\delta I \equiv I - 1$ via a statistical analysis. However, there is an important obstacle. In order to arrive at a sensitivity of cosmic ray anisotropy at the per-mille-level, the detector response has to be known at even {\it better} accuracy. This is difficult, if not impossible, to achieve via Monte-Carlo techniques. 

Instead of assuming a local detector acceptance and exposure, we can attempt a model-independent reconstruction by a simultaneous fit of these quantities together with the cosmic ray relative intensity. This method will come at the price of a lower statistical sensitivity to the cosmic ray anisotropy, but it compensates for the systematic uncertainty of the detector data. 

Unfortunately, there is an important limitation of this method. Note that events recorded at a fixed position $(\varphi, \theta)$ in the local coordinate system can only probe the cosmic ray flux along a constant declination $\delta(\varphi, \theta)$, {\it i.e.}, only variations of the flux with respect to right ascension $\alpha(t,\varphi, \theta)$ as the sidereal time increases. Hence, the expectation values (\ref{eq:mu}) are invariant under the simultaneous rescaling
\begin{eqnarray}
\label{eq:scaleI}
  I \to & I' &\equiv I/a(\delta)/b\,,\\\label{eq:scaleN}
  \mathcal{N}\to &{\mathcal{N}}'&\equiv \mathcal{N}bc
\,,\\\label{eq:scaleA}
  \mathcal{A} \to &{\mathcal{A}}'&\equiv{\mathcal{A}a(\delta(\varphi, \theta))}/{c}\,,\end{eqnarray}
where $a(\delta)$ is an arbitrary function of declination and the normalization factors $b$ and $c$ are defined such that $\int{\rm d}\Omega \mathcal{A}'(\Omega)=1$ and $\int{\rm d}\Omega \delta I'(\Omega)=0$ for the new values. In other words, the simultaneous reconstruction of the relative detector acceptance and isotropic background level does only allow us to reconstruct the relative intensity up to an azimuthally symmetric scaling function. 

A natural choice (see Appendix~\ref{AppendixA}) is that the anisotropy is normalized to $\int{\rm d}\alpha\delta I(\alpha,\delta)=0$ for all declinations $\delta$, consistent with the definition $\int{\rm d}\Omega \delta I(\alpha,\delta)=0$. This condition can also be formulated in terms of an expansion of the relative intensity into spherical harmonics, as pointed out by \cite{Iuppa:2013pg}. In general, the relative intensity can be decomposed as a sum over spherical harmonics $Y^{\ell m}$ as
\begin{equation}\label{eq:Ylm}
{\delta I}(\alpha,\delta) = \sum_{\ell\geq1}\sum_{m=-\ell}^\ell \widehat{a}_{\ell m} Y^{\ell m}(\pi/2-\delta,\alpha)\,.
\end{equation}
Our normalisation condition can then be expressed as the condition $\widehat{a}_{\ell 0}=0$ for all $\ell$. This projection significantly reduces the reconstruction of the low-$\ell$ multipole components of the anisotropy. Further details are provided in Appendix~\ref{AppendixA}.

Note that the true multipole moments $\widehat{a}_{\ell m}$ are an (infinite) superposition of the {\it pseudo} multipole moments ${a}_{\ell m}$, which are defined as in Eq.~(\ref{eq:Ylm}), but for the product of the relative intensity with the window function $w$ of the field of view. Provided that the window function is azimuthally symmetric, $w(\alpha,\delta) \simeq w(\delta)$, the true multipole moments $\widehat{a}_{\ell 0}$ are a linear superposition of pseudo multipole moments ${a}_{\ell' 0}$. In practice, we can hence use the normalization condition ${a}_{\ell 0} = 0$ for all $\ell$ to ensure $\widehat{a}_{\ell 0} = 0$ for all $\ell$. In terms of the binned relative intensity and window function, this is equivalent to the condition $\int{\rm d}\Omega w(\delta)Y^{\ell 0}(\pi/2-\delta,\alpha)\delta I(\alpha,\delta)=0$ for all $\ell$.

\section{Maximum-likelihood Method}\label{sec:MaxLH}

The number of cosmic rays expected from a solid angle $\Delta\Omega_i$ at the location $(\varphi_i,\theta_i)$ and a sidereal time interval $\Delta t_\tau$ with central value $t_\tau$ can be expressed as
\begin{equation}\label{eq:mubin}
  \mu_{\tau i} \simeq I_{\tau i}\mathcal{N}_\tau\mathcal{A}_{i}\,.
\end{equation}
Here, $\mathcal{N}_\tau\equiv\Delta t_\tau\mathcal{N}(t_\tau)$ is the expected number of isotropic background events in sidereal time bin $\tau$, $\mathcal{A}_i \equiv \Delta\Omega_i\mathcal{A}(\theta_i,\varphi_i)$ is the binned relative acceptance of the detector for angular element $i$, and $I_{\tau i}\equiv {I}({\bf R}(t_\tau){\bf n}'(\Omega_{i}))$ is the relative intensity observed in the local horizontal system during time bin $\tau$.  
 
To simplify calculations in what follows, we will assume that the solid angle bins in the local and celestial spheres are uniform, $\Delta\Omega_i=\Delta\Omega$, and that the sidereal time intervals are of equal size, $\Delta t_\tau = \Delta t$. We follow the conventions of \cite{Ahlers:2016njl} and use greek indices to indicate sidereal time bins, roman indices for bins in the local sky map, and fraktur indices for bins in the celestial sky map. Note that the relative intensity $I_\mathfrak{a} \equiv I({\bf n}(\Omega_{a}))$ and the local acceptance $\mathcal{A}_i \equiv \Delta\Omega\mathcal{A}({\bf n}'(\Omega_i))$ are assumed to be constant in the celestial and local coordinate systems, respectively, but can be transformed into the other coordinate system by the time-dependent rotation (\ref{eq:Rmatrix}). For instance, the quantity $\mathcal{A}_{\tau\mathfrak{a}}\equiv \Delta\Omega\mathcal{A}({\bf R}^T(t_\tau){\bf n}(\Omega_\mathfrak{a}))$ denotes the local detector acceptance for the solid angle $\Omega_\mathfrak{a}$ on the celestial sphere for the sidereal time step $\tau$.

Given the expectation values $\mu_{\tau i}$, the likelihood of observing a distribution of $n_{\tau i}$ cosmic ray events is given by the product of Poisson probabilities
\begin{equation}\label{eq:LH}
  \mathcal{L}(n|I,\mathcal{N},\mathcal{A}) =
  \prod_{\tau i}\frac{(\mu_{\tau i})^{n_{\tau i}}e^{-\mu_{\tau i}}}{n_{\tau i}!}\,.
\end{equation}
This likelihood can be maximized to provide estimators of the relative acceptance function $\mathcal{A}_i$ and the expected isotropic background count $\mathcal{N}_\tau$. In the absence of anisotropy, $I_\mathfrak{a}^{(0)}=1$, the maximum of the likelihood (\ref{eq:LH}) subject to the boundary condition $\sum_i\mathcal{A}_i=1$ is given by
\begin{align}\label{eq:Nnull}
  \mathcal{N}_\tau^{(0)} &=  {\sum_i n_{\tau i}}\,,\\\label{eq:Anull}
  {\mathcal{A}}_i^{(0)} &= \sum_\tau n_{\tau i}\Big/\sum_{\kappa j}n_{\kappa j}\,.
\end{align}
These estimators of the background rate and relative acceptance are analogous to the detector calibration methods used in time-scrambling~\citep{Alexandreas1993} or direct integration~\citep{Atkins:2003ep}. However, in the presence of anisotropies these first-order estimators can receive sizable corrections.

Allowing for the presence of anisotropy, we maximize the likelihood (\ref{eq:LH}) simultaneously in $I$, $\mathcal{N}$, and $\mathcal{A}$. The maximum
$(I^\star,\mathcal{N}^\star,\mathcal{A}^\star)$ of the likelihood (\ref{eq:LH}) must obey the implicit relations
\begin{align}\label{eq:Istar}
  {I}^\star_{\mathfrak{a}} &=
  \sum_{\tau} n_{\tau\mathfrak{a}}\Big/ \sum_{\kappa}\mathcal{A}^{\star}_{\kappa \mathfrak{a}}\mathcal{N}^{\star}_\kappa\,, \\\label{eq:Nstar}
  \mathcal{N}^{\star}_\tau &=
  \sum_{i}  n_{\tau i}\Big/\sum_{j}\mathcal{A}^{\star}_jI^\star_{\tau j}\,,\\\label{eq:Astar}
\mathcal{A}^{\star}_i&= \sum_\tau n_{\tau i}\Big/\sum_{\kappa}\mathcal{N}^{\star}_\kappa I^\star_{\kappa i}\,.
\end{align}
together with the normalization conditions $\sum_\mathfrak{a}w_\mathfrak{a}Y^{\ell 0}_\mathfrak{a}\delta I^\star_\mathfrak{a} =0$ and $\sum_i \mathcal{A}^{\star}_i =1$. Equations~\eqref{eq:Istar}, \eqref{eq:Nstar}, and \eqref{eq:Astar} correspond to a nonlinear set of equations that cannot be solved in an explicit form. However, one can approach the best-fit solution iteratively by an algorithm introduced in \cite{Ahlers:2016njl} and outlined in Appendix~\ref{AppendixB}. 

In the case of limited statistics, which is typical for cosmic ray observations above the knee, the iterative method by \cite{Ahlers:2016njl} needs to be adapted to increase the stability of the numerical reconstruction. The simplest way is by smoothing the event distribution $n_{\tau i}$ with a Gaussian beam with an appropriate angular size. This procedure will only affect the small-scale anisotropy that is present in the data, but undistinguishable from the noise introduced by Poisson fluctuations. 

Instead of smoothing the original event map to account for the limited statistics in cosmic ray data above the ankle, it is also possible to adapt the maximum-likelihood method to account for a smoothing scale in the relative intensity. This can be done by an expansion of the anisotropy into spherical harmonics (\ref{eq:Ylm}) that is truncated at a maximum moment $\ell_{\rm max}$. We discuss the case of a general truncation scale $\ell_{\rm max}$ in Appendix~\ref{AppendixC} and concentrate here on the dipole anisotropy, $\ell_{\rm max}=1$. In this case, it is convenient to work with the expansion
\begin{equation}\label{eq:Idipole}
\delta I_{\rm dipole}(\alpha,\delta) = d_xx(\alpha,\delta) + d_yy(\alpha,\delta)\,,
\end{equation}
where $x(\alpha,\delta) = \cos\alpha\cos\delta$ and $y(\alpha,\delta)= \sin\alpha\cos\delta$. These basis functions correspond to the projection of the unit vector ${\bf n}$ into the equatorial plane. The relation to spherical harmonics is $x=\sqrt{2\pi/3}(Y^{1-1}-Y^{11})$ and $y=i\sqrt{2\pi/3}(Y^{1-1}+Y^{11})$ and therefore 
$a_{1-1}=-a^*_{11}=\sqrt{2\pi/3}(d_x+id_y)$. Note that the third component of ${\bf n}$ perpendicular to the equatorial plane is proportional to $Y^{10}$, which is not accessible by this data-driven method, as explained in section~\ref{sec:Observation}. The dipole (\ref{eq:Idipole}) automatically satisfies the normalization condition $\sum_\mathfrak{a} \delta I_\mathfrak{a} =0$.

With this ansatz for the relative intensity, the maximum-likelihood solution $(d^\star_x,d^\star_y,\mathcal{N}^\star,\mathcal{A}^\star)$ for a $d^\star_x\ll1$ and  $d^\star_y\ll1$ is given by Eqs.~(\ref{eq:Nstar}) and (\ref{eq:Astar}) together with the simple matrix equation (see Appendix~\ref{AppendixB} for details)
\begin{multline}\label{eq:dxdystar}
\sum_{\tau i}n_{\tau i}\begin{pmatrix}x^2_{\tau i}&x_{\tau i}y_{\tau i}\\x_{\tau i}y_{\tau i}&y^2_{\tau i}\end{pmatrix}\begin{pmatrix}d^\star_x\\d^\star_y\end{pmatrix}\\\simeq\sum_{\tau i}\begin{pmatrix}(n_{\tau i}-\mathcal{N}^\star_\tau\mathcal{A}^\star_i)x_{\tau i}\\(n_{\tau i}-\mathcal{N}^\star_\tau\mathcal{A}^\star_i)y_{\tau i} \end{pmatrix}
\,.
\end{multline}
Here, we again make use of the notation $x_{\tau i}\equiv x({\bf R}(t_\tau){\bf n}'(\Omega_{i}))$, etc. As before, the nonlinear system of equations (\ref{eq:Nstar}), (\ref{eq:Astar}), and (\ref{eq:dxdystar}) can only be solved via an iterative reconstruction method outlined in Appendix~\ref{AppendixB}. 

Another advantage of the likelihood-based dipole reconstruction method is the simplicity of estimating the significance of the observation. The maximum-likelihood ratio between the best-fit dipole anisotropy and the null hypothesis, $I=1$, defines the maximum-likelihood test statistic
\begin{equation}
\lambda = 2\ln\frac{\mathcal{L}(n|d_x^\star,d_y^\star,\mathcal{N}^\star_\tau,\mathcal{A}^\star_i)}{\mathcal{L}(n|0,0,\mathcal{N}^{(0)}_\tau,\mathcal{A}^{(0)}_i)}\,.
\end{equation}
According to \cite{Wilks:1938dza}, data following the null hypothesis have a distribution in $\lambda$ that follows a two-dimensional $\chi^2$-distribution. The $p$-value of the observed data, {\it i.e.}, the probability of a false positive identification of the dipole anisotropy, is simply given by $p= e^{-\lambda/2}$. 

We can also use the maximum-likelihood (\ref{eq:LH}) to estimate the parameter uncertainties, $\sigma_{x/y}$, of the dipole amplitudes $d^\star_{x/y}$. The derivation for the covariance matrix for general $\ell_{\rm max}$ is discussed in Appendix~\ref{AppendixC}. For the case of the dipole anisotropy it can be well approximated as
\begin{equation}\label{eq:sigma}
\sigma^{-2}_x \simeq \sum_{\tau i}n_{\tau i} (x_{\tau i})^2 - \sum_\tau \frac{(\mathcal{N}^\star_\tau)^2}{\sum_i n_{\tau i}}\bigg(\sum_j \mathcal{A}_j^\star x_{\tau j}\bigg)^2\,,
\end{equation}
with an analogous equation for the uncertainty $\sigma_y$ of the second component $d_y$. The first term of  expression (\ref{eq:sigma}) is approximately $N_{\rm tot}/2$, where $N_{\rm tot}$ is the total event number. This corresponds to the naive first-order approximation $\sqrt{2/N_{\rm tot}}$ of the uncertainty. However, the second term increases the error in the dipole reconstruction. This accounts for the fact that the statistical power of the data is also used to separately determine the background rate. As we will see in the next section, this will lead to a weaker significance of the Auger dipole reconstruction, compared to the original analysis in~\cite{Aab:2017tyv}.

\section{Analysis of Auger Data}\label{sec:Auger}

\begin{figure}[t]\centering
\includegraphics[width=\linewidth,viewport= 0 0 500 275]{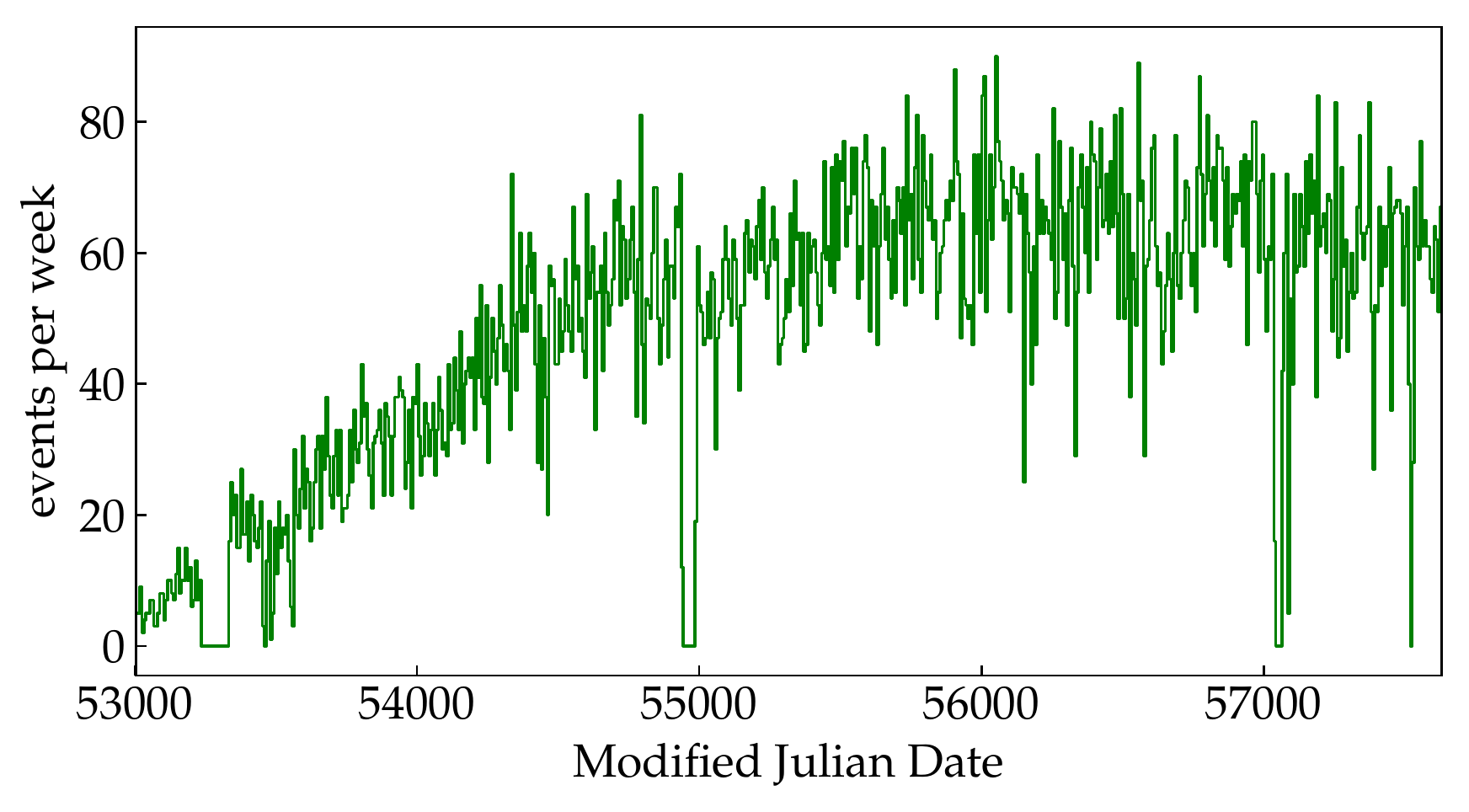}\\
\includegraphics[width=\linewidth,viewport= 8 0 508 275]{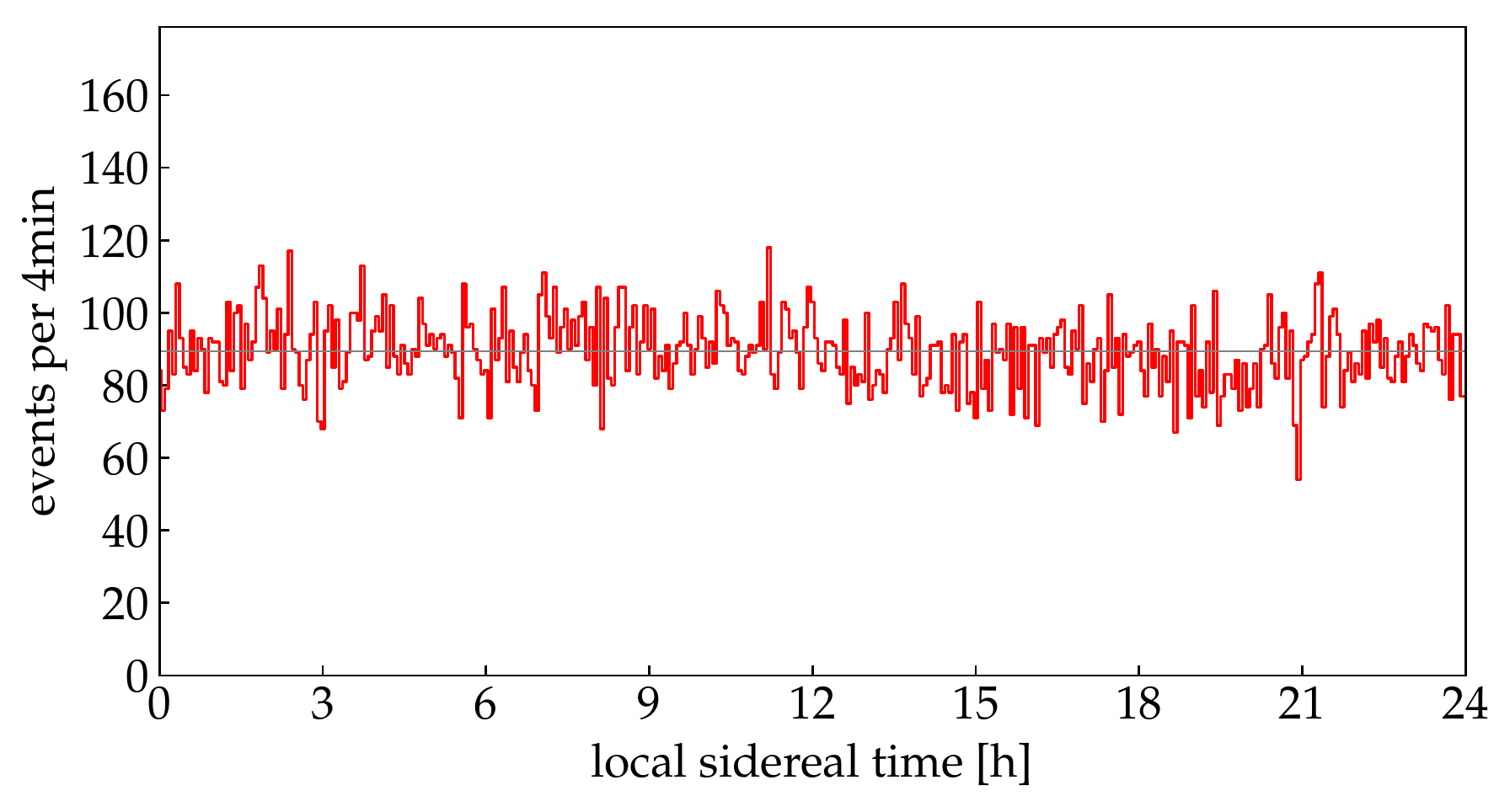}
\caption[]{Arrival time of events with $E>8$~EeV in terms of modified Julian days (top) and sidereal time (bottom). The horizontal thin line in the lower plot indicates the mean number of events per 20 minutes.}\label{fig:Augerdata}
\end{figure}

We will now apply the previously discussed methods to the Auger data at energies above 8~EeV. The Pierre Auger Observatory~\citep{ThePierreAuger:2015rma} is located near the city of Malarg\"ue, Argentina, at a geographic latitude of $\Phi\simeq35.2^\circ$S and longitude $\Lambda\simeq69.5^\circ$W. The 32187 cosmic ray events used in this analysis have been recorded over a 12 year period from 2004 January to 2016 August. The arrival times in terms of Median Julian Days (MJDs) are shown in the top panel of Fig.~\ref{fig:Augerdata}. The strong rise of the overall event rate during the first $\sim1500$ MJDs can be attributed to the growth of the detector while it was already taking data. This changing detector configuration can be compensated by the maximum-likelihood reconstruction as long as the relative acceptance (averaged over many sidereal days) remains independent of sidereal time. The data distribution over sidereal time is shown in the bottom panel of Fig.~\ref{fig:Augerdata}.

The cosmic ray data cover a wide range of reconstructed zenith angles up to $80^\circ$. The integrated field of view corresponds to a wide declination range of $-90^\circ\leq\delta\leq44.8^\circ$ that provides excellent condition for the reconstruction of large-scale anisotropies. The Pierre Auger Collaboration uses two different reconstruction methods for data with zenith angles $\theta<60^\circ$ and $60^\circ\leq\theta\leq80^\circ$~\cite{Aab:2017tyv}. After reconstruction of the zenith angles via Eq.~(\ref{eq:Rmatrix}) using the reported arrival time, right ascension angle $\alpha$, and declination $\delta$ in the official data release, these two data sets are clearly visible in the local distribution of events. We also observe a small mutual tilt ($\Delta\theta\lesssim1^\circ$) of the corresponding event distributions. However, this has little effect on the following analysis, due to the model-independent reconstruction of the detector response.

\begin{figure*}[p]\centering
\includegraphics[width=0.8\linewidth,viewport= 30 0 640 370,clip=true]{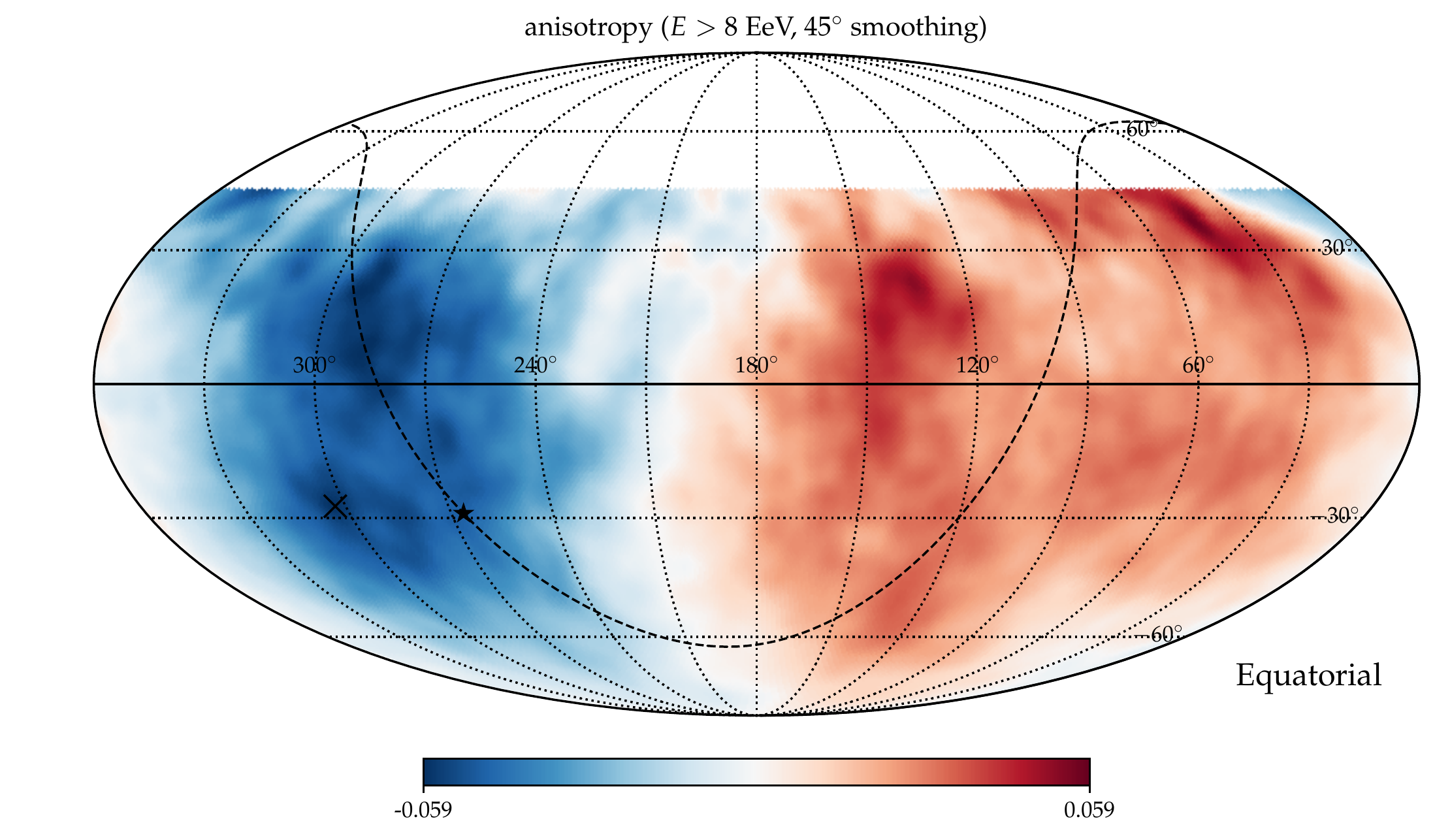}\\
\includegraphics[width=0.8\linewidth,viewport= 30 0 640 370,clip=true]{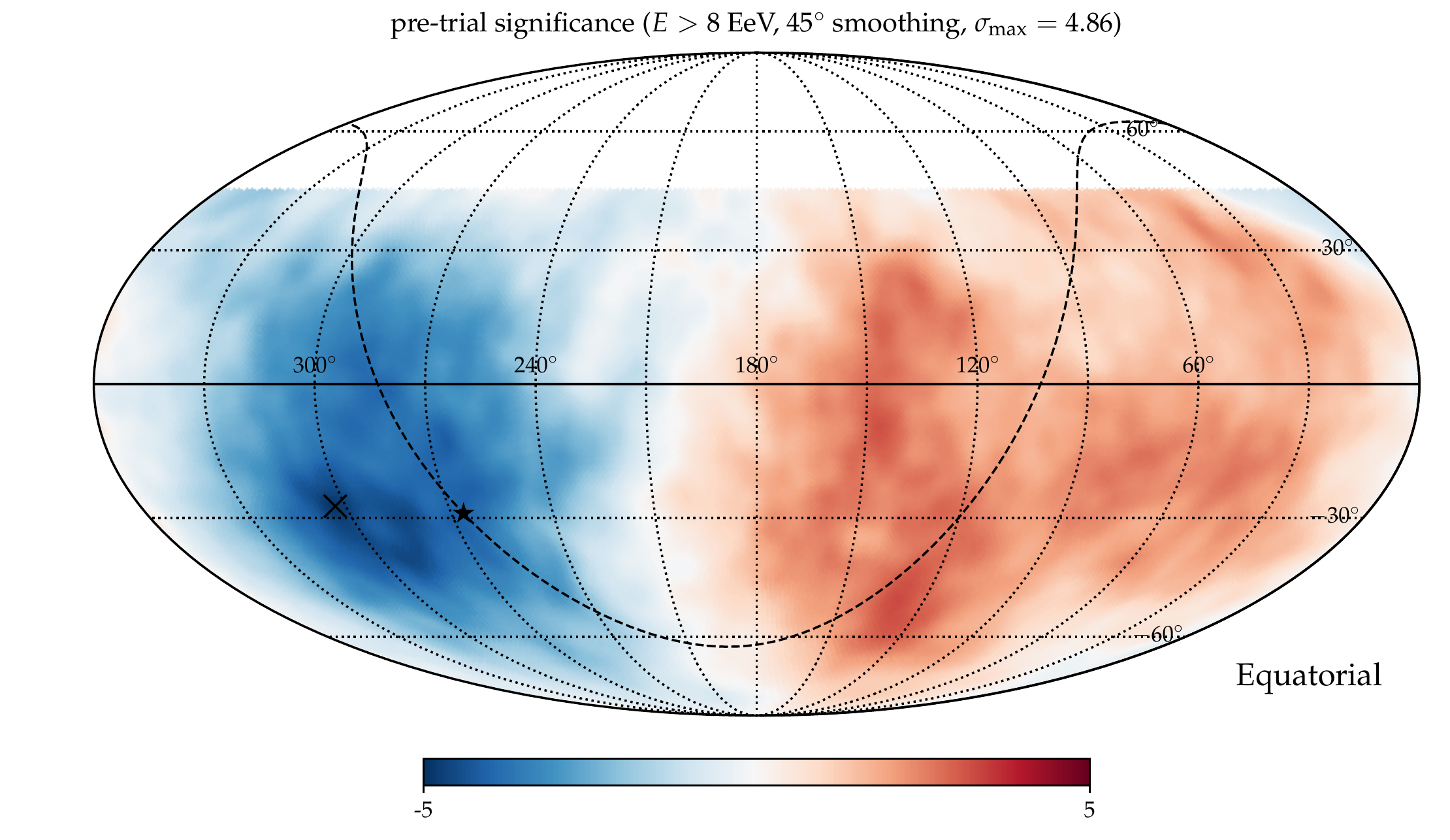}
\caption[]{Mollweide projection of the reconstructed anisotropy (top) and significance (bottom) of the Auger data above 8~EeV in the Equatorial coordinate system. We smooth the anisotropy and significance over a radius of $45^\circ$ following the weighting procedure described in the main text. The dashed line and star indicate the position of the Galactic plane and center, respectively. The black cross in both maps indicates the location of the highest pre-trial significance of the smoothed map of $4.86\sigma$ corresponding to a local deficit.}\label{fig:AugerLarge}
\end{figure*}

\subsection{Large-scale Anisotropy}

We first apply the dipole reconstruction method discussed in Section~\ref{sec:MaxLH}. The iterative method (see Appendix~\ref{AppendixB}) converges after a few iterations and we terminate the reconstruction after 10 steps. The two dipole orientations have the best-fit values $d_x = (-1.2\pm1.3)\%$ and $d_y = (5.0\pm1.3)\%$, where the errors indicate the $1\sigma$ statistical uncertainties. In terms of dipole amplitude in the equatorial plane we have $A_\perp=(5.3\pm1.3)$\% and a right ascension angle of $103^\circ\pm15^\circ$. The maximum-likelihood value of the best fit is $\lambda\simeq 14.77$ which translates into a $p$-value of $6.2\times10^{-4}$ or a significance of $3.4\sigma$ assuming the applicability of the Wilks theorem~\cite{Wilks:1938dza}. Assuming that the likelihood function in $d_x$ and $d_y$ follows a two-dimensional Gaussian distribution, we arrive at a similar significance of $3.5\sigma$.

In preparation for the small-scale anisotropy analysis in the following subsection, we also apply the full likelihood-based anisotropy reconstruction introduced by \cite{Ahlers:2016njl}. To increase the stability of the iterative reconstruction, we smooth the data with a Gaussian symmetric beam with full width half maximum of $4^\circ$. Similar to the case of the dipole reconstruction, the iterative method converges after a few iterations and is terminated after 10 steps. For comparison with the original result of \cite{Aab:2017tyv} we smooth the resulting anisotropy over a radius of $45^\circ$. This is done by first rebinning expectation values and event numbers into sliding bins centered around a position $\Omega_\mathfrak{a}$ in the equatorial coordinate system as
\begin{align}\label{eq:rebinning}
  \widetilde{n}_{\mathfrak{a}} &= \sum_{\mathfrak{b}\in\mathcal{D}_{\mathfrak{a}}}\sum_{\tau}n_{\tau \mathfrak{b}}\,,\\
  \widetilde{\mu}_{\mathfrak{a}} &=  \sum_{\mathfrak{b}\in\mathcal{D}_{\mathfrak{a}}}\sum_{\tau}\mathcal{A}^\star_{\tau\mathfrak{b}}\mathcal{N}^\star_\tau I^\star_{\mathfrak{b}}\,,\\
  \widetilde{\mu}^{\,\rm bg}_{\mathfrak{a}} &= \sum_{\mathfrak{b}\in\mathcal{D}_{\mathfrak{a}}}\sum_{\tau}\mathcal{A}^\star_{\tau\mathfrak{b}}\mathcal{N}^\star_\tau I^{\,\rm bg}_{\mathfrak{b}}\,.
\end{align}
where $\mathcal{D}_\mathfrak{a}$ denotes the set of data bins within $45^\circ$ of the location $\Omega_\mathfrak{a}$. The isotropic background level is simply $I^{\,\rm bg}=1$. With this definition we can define the smoothed anisotropy as 
\begin{equation}\label{eq:rebinnedI}
\delta\widetilde{I}_\mathfrak{a} = \widetilde{\mu}_\mathfrak{a}/\widetilde{\mu}^{\,\rm bg}_\mathfrak{a} - 1
\end{equation}
The top panel of Figure~\ref{fig:AugerLarge} shows the result of this smoothing procedure. The anisotropy is shown as a Mollweide projection in the equatorial coordinate system with excesses and deficits indicated by red and blue colors, respectively. The dashed line and star indicate the position of the Galactic plane and the center. The anisotropy agrees qualitatively with the result of \cite{Aab:2017tyv} (their Fig.~2). Note, however, that the likelihood-based reconstruction is not sensitive to the $a_{\ell0}$ coefficients of the spherical harmonic expansion.

\begin{figure*}[t]\centering
\includegraphics[width=0.5\linewidth,viewport= 30 0 640 370,clip=true]{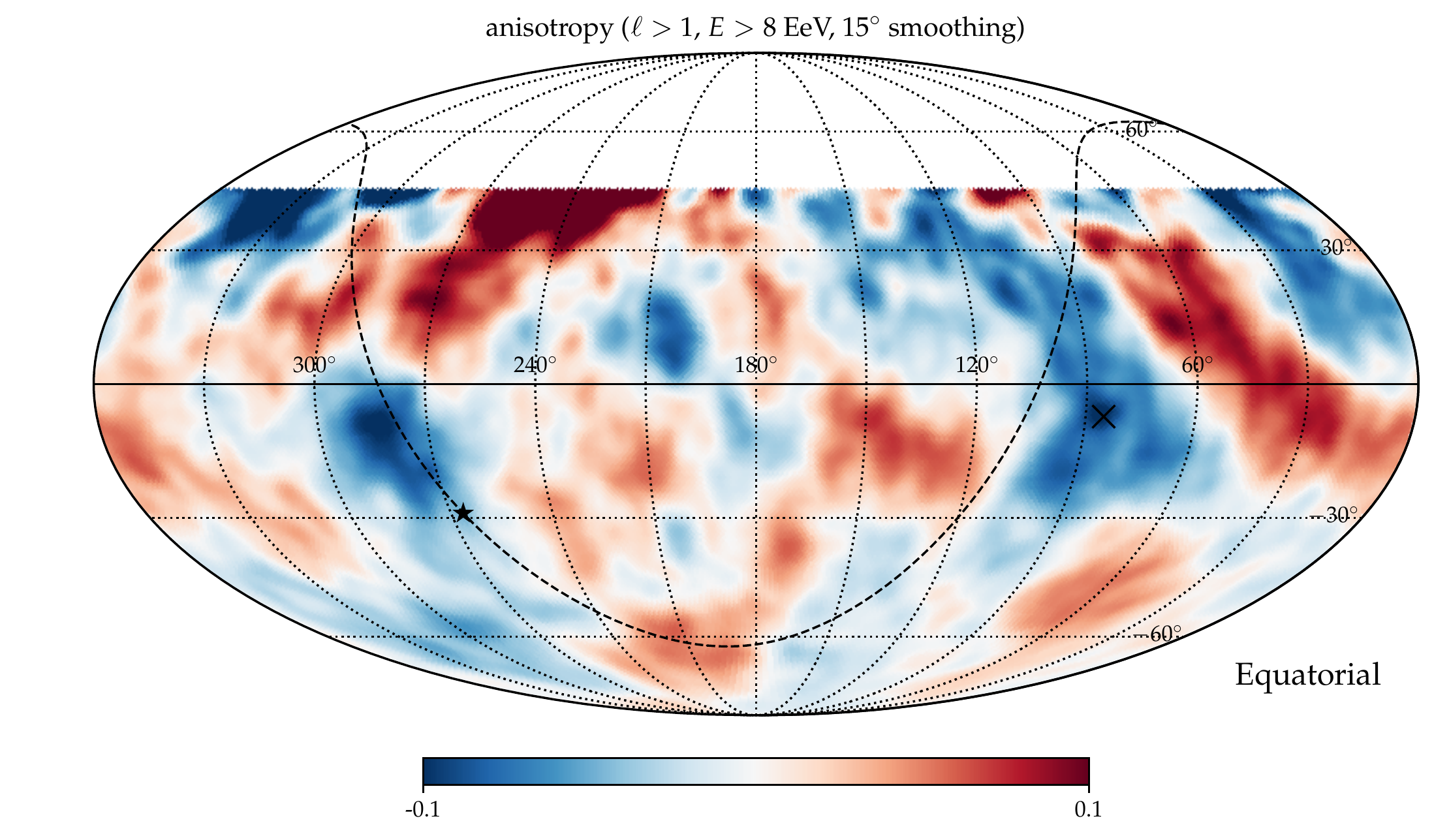}\hfill\includegraphics[width=0.5\linewidth,viewport= 30 0 640 370,clip=true]{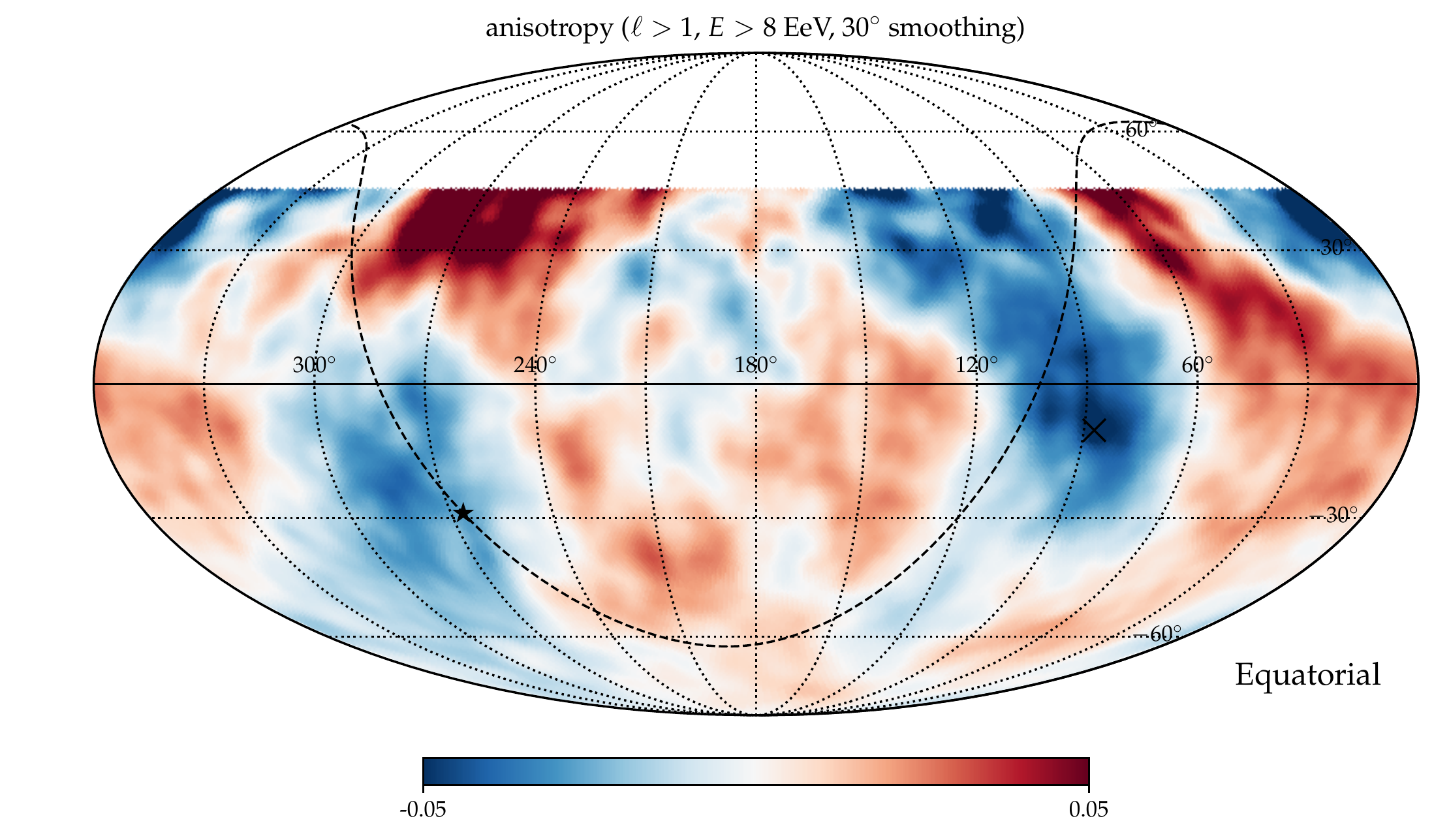}\\
\includegraphics[width=0.5\linewidth,viewport= 30 0 640 370,clip=true]{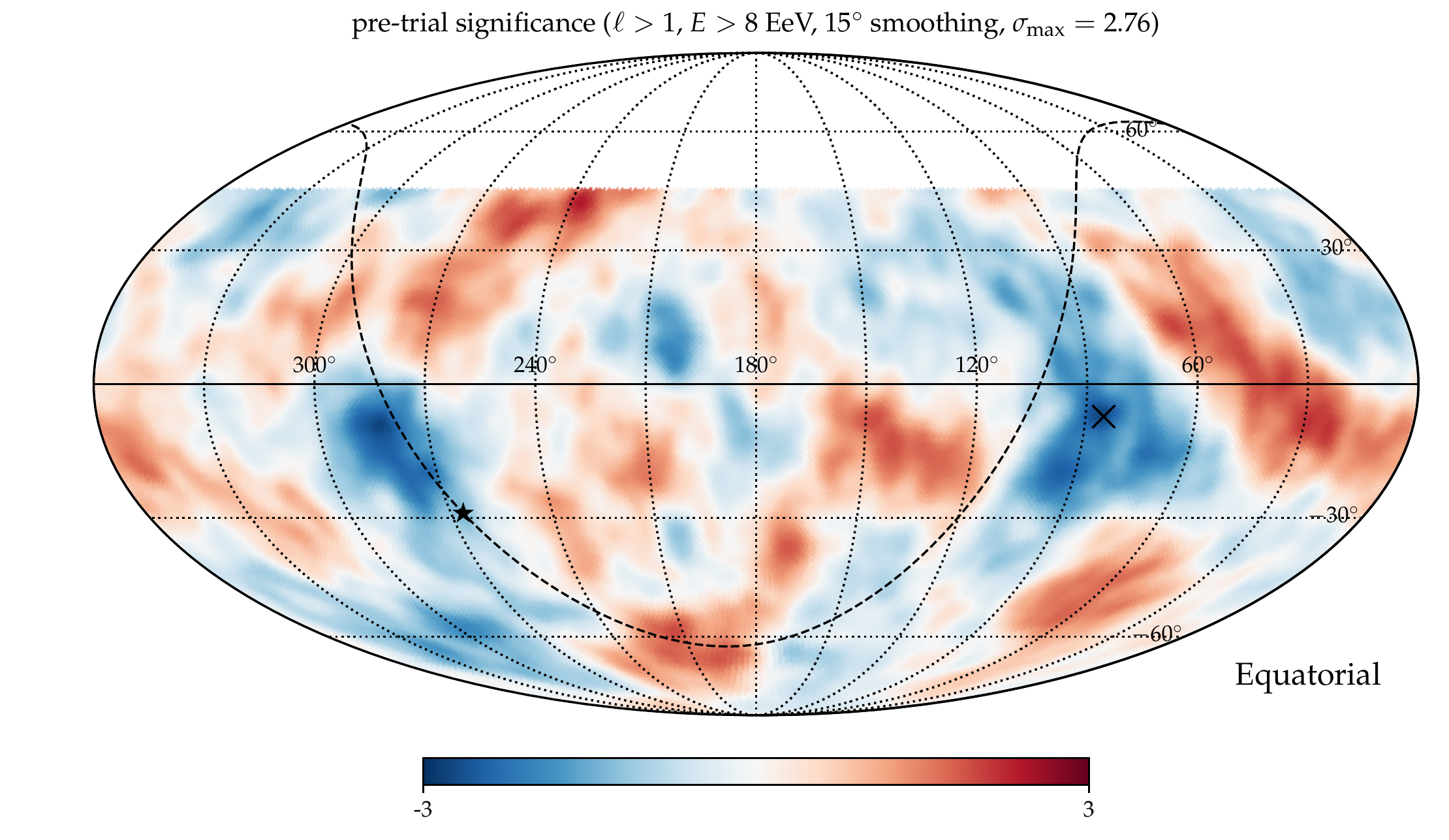}\hfill\includegraphics[width=0.5\linewidth,viewport= 30 0 640 370,clip=true]{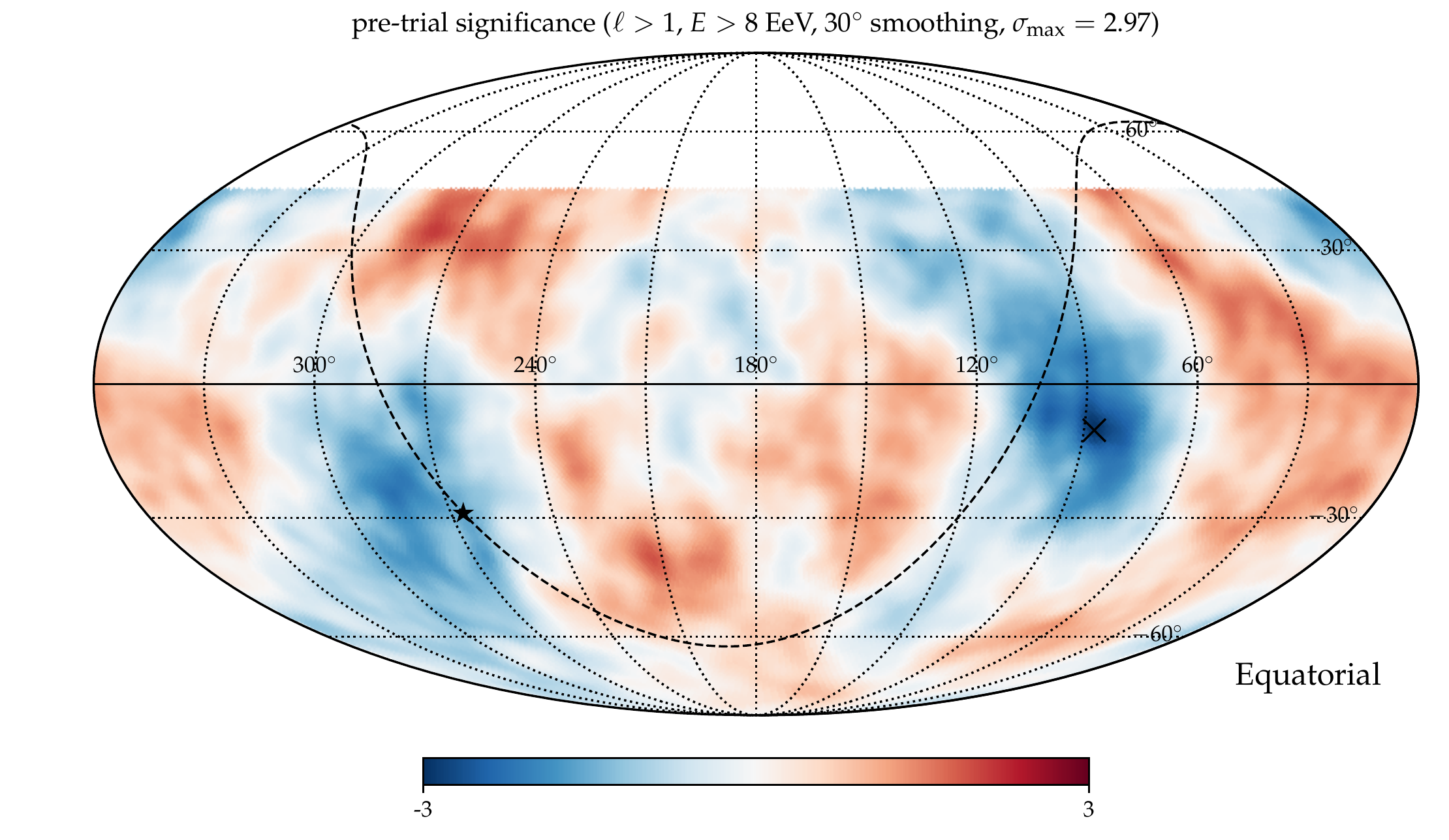}
\caption[]{Same as Fig.~\ref{fig:AugerLarge}, but now showing the Mollweide projection of the anisotropy (top) and significance (bottom) after removal of the dipole anisotropy and smoothing over a radius of $15^\circ$ (left) and $30^\circ$ (right). Note that the color scale of the pre-trial significance is adjusted to $\pm 3\sigma$. The maximum pre-trial significance is $2.76\sigma$ ($2.97\sigma$) for the $15^\circ$ ($30^\circ$) smoothing, which is reached at the position indicated by black crosses in the map.}\label{fig:AugerSmall}
\end{figure*}

With the rebinned data and expectation values of Eqs.~(\ref{eq:rebinning}) we can also define a significance map as
\begin{equation}\label{eq:rebinnedsig}
  \widetilde{S}_{\mathfrak{a}} \equiv \sqrt{2\left(-\widetilde{\mu}_{\mathfrak{a}}+\widetilde{\mu}^{\,\rm bg}_{\mathfrak{a}} + \widetilde{n}_\mathfrak{a}\log(1+\delta\widetilde{I}_\mathfrak{a})\right)}\,.
\end{equation}
This expression represents the statistical weight of the anisotropy $I_\mathfrak{a}^\star-I_\mathfrak{a}^{\,\rm bg}$ in each celestial (sliding) bin $\mathfrak{a}$. For sufficiently small smoothing scales, $(\widetilde{S}_\mathfrak{a})^2$ can be interpreted as the bin-by-bin maximum-likelihood ratio of the hypothesis $I^\star_\mathfrak{a}$ compared to the null hypothesis $I^{\,\rm bg}_\mathfrak{a}$. Again, according to \cite{Wilks:1938dza}, the test statistic of data under the null hypothesis is following a one-dimensional $\chi^2$-distribution and, in that case, $\widetilde{S}_{\mathfrak{a}}$ corresponds to the significance in units of Gaussian $\sigma$. Note, however, that this is only the pre-trial significance, which accounts for trials factors. The post-trial $p$-value can be approximated as $p_{\rm post} \simeq 1-(1-p_{\rm pre})^{N_{\rm trial}}$ with effective number of trials $N_{\rm trial}$. We will approximate this trial factor in the following by the ratio 
\begin{equation}
N_{\rm trial} \simeq \Delta\Omega_{\rm FOV}/\Delta\Omega_{\rm bin}\,,
\end{equation}
where $\Delta\Omega_{\rm FOV}$ is the size of the observatory's time-integrated field of view and $\Delta\Omega_{\rm bin}$ is the effective bin size according to the smoothing scale.
 
The lower panel of Figure~\ref{fig:AugerLarge} shows the smoothed pre-trial significance for the $45^\circ$ smoothed anisotropy in the top panel. Here, we follow the convention and indicate the significance of deficits (blue regions in the top panel) with negative significance values. The black crosses in both maps indicate the positions with the highest statistical weight corresponding to a deficit with a pre-trial significance of $4.86\sigma$. We can estimate a trial factor of $N_{\rm trial}\simeq5.8$ with a corresponding post-trial significance of $4.5\sigma$. 

\subsection{Small-scale Anisotropy}

We now turn to the question of the presence of small-scale anisotropy. Using the best-fit dipole anisotropy (\ref{eq:Idipole}) from the previous section, we can define a new background anisotropy as $I^{\,\rm bg} = 1 + \delta I_{\rm dipole}$ and calculate the smoothed anisotropy and significance as in Eqs.~(\ref{eq:rebinnedI}) and (\ref{eq:rebinnedsig}). The smoothed residual anisotropy and corresponding significance maps are shown in Fig.~\ref{fig:AugerSmall}. We assume smaller smoothing scales of $15^\circ$ (left panels) and $30^\circ$ (right panels) compared to the large $45^\circ$ smoothing scale used in Fig.~\ref{fig:AugerLarge}. We cannot find evidence for small-scale anisotropies in the residual anisotropy maps. The pre-trial significance shown in the lower panels stays below $3\sigma$ at any location in the equatorial coordinate system. The maximal pre-trial significance for the $15^\circ$ ($30^\circ$) smoothing is $2.76\sigma$ ($2.97\sigma$) and corresponds to a local deficit indicated by a black cross. We can estimate a trial factor of $N_{\rm trial}\simeq50$ ($12.7$) resulting in a post-trial significance of $1.1\sigma$ ($2.1\sigma$).

\begin{figure}[t]
\includegraphics[width=\linewidth]{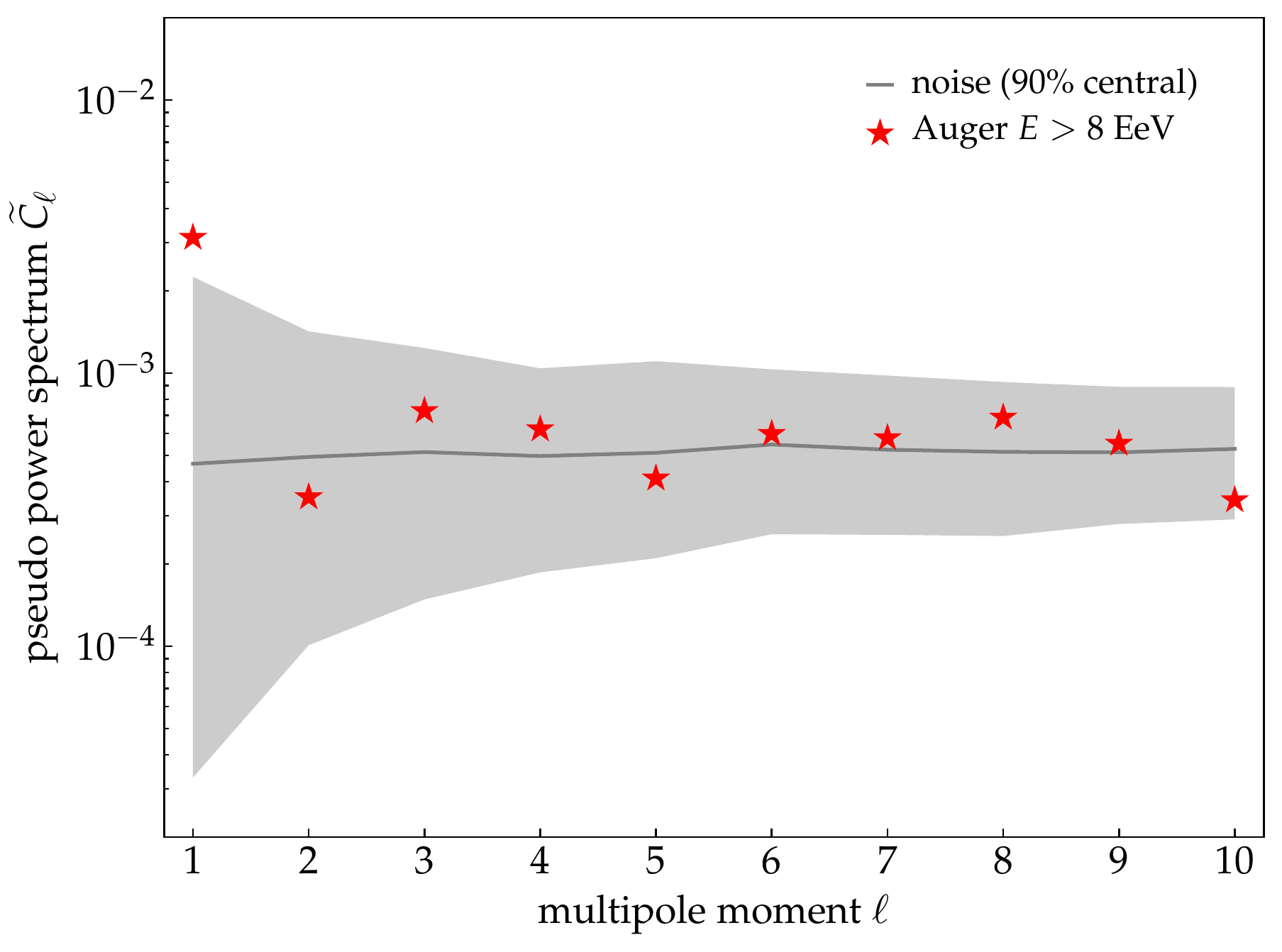}
\caption[]{The pseudo power spectrum of the Auger data above 8~EeV. The gray shaded band shows the median pseudo power spectrum and central 90\% range of 1000 background simulations with no anisotropy, $\delta I =0$. For the simulation of the background data we use the best-fit background rate $\mathcal{N}^\star$ and relative detector acceptance $\mathcal{A}^\star$ from the analysis of the actual data. The anisotropy of each background simulation is then reconstructed with the same iterative likelihood method.}\label{fig:powerspectrum}
\end{figure}

An alternative method to quantify the presence of small-scale anisotropies in the data is by the pseudo power spectrum defined as
\begin{equation}
\widetilde{C}_\ell = \frac{1}{2\ell+1}\sum_{m=-\ell}^\ell|\widetilde{a}_{\ell m}|^2\,,
\end{equation}
where $\widetilde{a}_{\ell m}$ are the coefficients of the anisotropy expansion into spherical harmonics under the assumption of a full sky coverage. Note, that due to the reduced field of view ($-90^\circ\leq\delta\leq44.8^\circ$) and the level of bin-by-bin fluctuations, the pseudo coefficients $\widetilde{a}_{\ell m}$ are different from the true coefficients $a^\star_{\ell m}$. Nevertheless, we can use the pseudo power $\widetilde{C}_\ell$ as a test statistic to probe significant deviations from a background hypothesis. The pseudo power spectrum of the reconstructed anisotropy map for $\ell\leq10$ (corresponding to angular scales larger than about $20^\circ$) is shown as the red data in Fig.~\ref{fig:powerspectrum}. 

We can now compare the observed pseudo power spectrum to the distribution of power spectra generated by mock data following the null hypothesis, $\delta I=0$. For a realistic detector description, we use the best-fit background rate $\mathcal{N}^\star$ and relative detector acceptance $\mathcal{A}^\star$ from the analysis of the actual data above 8~EeV, but generate the mock data following an isotropic distribution, $I=1$. The resulting data are then analyzed following the same iterative method (Appendix~\ref{AppendixB}) that we applied to the true data and the resulting best-fit anisotropy is analyzed in terms of its pseudo power spectrum. The result of this background simulation is shown by the median power spectrum and central 90\% range of 1000 mock data samples in Fig.~\ref{fig:powerspectrum}. As expected, the dipole anisotropy shows an excess with respect to the background level, but all other moments are consistent within background variations. In summary, there are no statistically significant small-scale anisotropies present in the cosmic ray data above 8~EeV.

\section{Conclusion}\label{sec:Conclusion}

We have analyzed data from the Pierre Auger Observatory for the presence of anisotropies in the arrival directions of cosmic rays above 8~EeV. Our analysis is based on a maximum-likelihood method, that simultaneously fit the relative detector acceptance and cosmic ray background rate and that is therefore independent of a precise modeling of the detector. With our method we independently derive a cosmic ray dipole anisotropy in the equatorial plane with an amplitude $A_\perp=(5.3\pm1.3)$\% and right ascension angle of $103^\circ\pm15^\circ$, consistent with the official result of the Pierre Auger Collaboration. 

The pre-trial significance of the dipole anisotropy is at the level of $3.4\sigma$, which is much lower than the official result of $5.6\sigma$ (pre-trial). The reason for this difference is related to the fact that the analysis of the Pierre Auger Collaboration relies on a detector model with an estimated systematic uncertainty below the $1\%$ level. Our likelihood-based reconstruction is less sensitive, but not limited by detector systematics related to the relative detector acceptance and background rate.
 
Our reconstruction method also allows us to study the presence of medium- and small-scale anisotropies in the data. We have analyzed the residual relative intensity of cosmic ray arrival directions after subtraction of the best-fit dipole anisotropy. No statistically significant medium-scale anisotropies larger than a smoothing radius of $15^\circ$ or $30^\circ$ are visible in the data. We have also analyzed the presence of medium- and small-scale anisotropies by comparing the pseudo power spectrum of the Auger anisotropy with simulated background maps. Also, in this case we cannot identify statistically significant anisotropies other than a dipole.

The analysis method discussed in this paper is well-suited for the future study of cosmic ray anisotropies below the ankle, where surface detectors are not fully efficient. Previously, the Pierre Auger Collaboration has analyzed the large-scale anisotropy in their low-energy data with the ``east-west method''~\citep{Bonino:2011nx,Abreu:2011ve}. This method also allows us to compensate for the systematic uncertainties of the detector. However, this outdated analysis technique projects the cosmic ray anisotropy into right-ascension bins and is therefore not capable of reconstructing a faithful two-dimensional anisotropy, as provided by the maximum-likelihood technique. 
 
\begin{acknowledgements}
{\it Acknowledgements.} I would like to thank the members of the Pierre Auger Collaboration for providing the arrival time of their cosmic ray data above $8$~EeV, which was needed for the likelihood-based reconstruction discussed in this paper. In particular, I would like thank Jonathan Biteau, Olivier Deligny, and Karl-Heinz Kampert for their comments regarding the official UHE CR anisotropy analysis and their feedback on my results. This work is supported by Danmarks Grundforskningsfond (project no.~1041811001) and by VILLUM FONDEN (project no.~18994).
\end{acknowledgements}

\appendix

\section{Scaling Invariance}\label{AppendixA}

The symmetry transformations in Eqs.~(\ref{eq:scaleI}), (\ref{eq:scaleN}), and (\ref{eq:scaleA}) imply that the relative intensity $I$ can only be determined up to a multiplicative function $a(\delta)$. Solving Eq.~(\ref{eq:scaleI}) for the rescaled anistropy gives
\begin{equation}\label{eq:dIp}
\delta I'(\alpha,\delta) = \frac{1+\delta I(\alpha,\delta)}{a(\delta)b}-1\,,
\end{equation}
where the factor $b$ is fixed by the condition $\int{\rm d}\Omega \delta I'(\alpha,\delta) =0$ and has the explicit form
\begin{equation}
b = \frac{1}{4\pi}\int{\rm d}\Omega\frac{1+\delta I(\alpha,\delta)}{a(\delta)}\,.
\end{equation}
The family of solutions implied by the transformation (\ref{eq:dIp}) include the true anisotropy $\delta \widehat{I}(\alpha,\delta)$. In order to extract a particular member of the family we can fix a ``scaling condition'' that is consistent with the normalization $\int {\rm d}\Omega \delta I(\alpha,\delta) =0$. One possibility is to fix the anisotropy by the condition
\begin{equation}\label{eq:dIfix}
\int{\rm d}\alpha \delta I^{\rm fix}(\alpha,\delta) = 0\,.
\end{equation}
This is a natural choice in the sense that it trivially implies the former condition. Note that this condition singles out a unique solution, as one can infer from the azimuthally averaged Eq.~(\ref{eq:dIp}): two solutions, $\delta I'^{\rm fix}$ and $\delta I^{\rm fix}$, imply $a(\delta)b=1$ and are therefore identical. On the other hand, if we had found a solution $\delta I$ that does not obey condition (\ref{eq:dIfix}), we can find the fixed solution $\delta I^{\rm fix}$ by the transformation (\ref{eq:dIp}) using the scaling function
\begin{equation}\label{eq:fix}
a(\delta)b = \frac{1}{2\pi}\int{\rm d}\alpha(1+\delta I(\alpha,\delta))\,.
\end{equation}

Now, the true anisotropy, $\delta \widehat{I}$, can be decomposed into spherical harmonics as in Eq.~(\ref{eq:Ylm}) with coefficients $\widehat{a}_{\ell m}$. Applying the scaling condition (\ref{eq:dIfix}) to $\delta \widehat{I}$ implies the transformation (\ref{eq:dIp}) with scaling function
\begin{equation}
a(\delta)b = 1+\sum_{\ell\geq1} \widehat{a}_{\ell 0} Y^{\ell 0}(\pi/2-\delta,\alpha)\,.
\end{equation}
This expression is identical to the azimuthally averaged relative intensity, $\langle\widehat{I}\,\rangle(\delta)$. From Eq.~(\ref{eq:dIp}) we therefore arrive at the (unique) solution
\begin{equation}
\delta I^{\rm fix}(\alpha,\delta) = \frac{1}{\langle\widehat{I}\,\rangle(\delta)}\sum_\ell\sum_{m\neq 0} \widehat{a}_{\ell m} Y^{\ell m}(\pi/2-\delta,\alpha)\,.
\end{equation}
Since the azimuthally averaged relative intensity is dominated by the monopole, $\langle\widehat{I}\,\rangle(\delta)\simeq 1$, this expression approximates the spherical harmonic expansion of the true anisotropy, $\delta\widehat{I}$, except for azimuthally symmetric components with $m=0$.

\section{Iterative Method}\label{AppendixB}

The set of nonlinear equations (\ref{eq:Istar}), (\ref{eq:Nstar}), and (\ref{eq:Astar}) together with the normalization condition $\sum_i\mathcal{A}^\star_i=1$ and gauge fixing $\widehat{a}_{\ell 0}=0$ can be solved via the following iterative method~\citep{Ahlers:2016njl}:
\begin{enumerate}[\rm (i)]
\item Initialize at the maximum of the null hypothesis, $({I}^{(0)},\mathcal{N}^{(0)},\mathcal{A}^{(0)})$. 
\item Evaluate ${I}^{(n+1)}$ by inserting $({I}^{(n)},\mathcal{N}^{(n)},\mathcal{A}^{(n)})$ into the right side of Eq.~(\ref{eq:Istar}).
\item Remove the $m=0$ (pseudo) multipole moments of $\delta{I}^{(n+1)}$, {i.e.}, in the equatorial coordinate system $\sum_\mathfrak{a}w_\mathfrak{a}Y^{\ell 0}_\mathfrak{a}\delta {I}_\mathfrak{a}^{(n+1)} \to 0$.
\item Evaluate $\mathcal{N}^{(n+1)}$ by inserting $({I}^{(n+1)},\mathcal{N}^{(n)},\mathcal{A}^{(n)})$ into the right side of Eq.~(\ref{eq:Nstar}).
\item Evaluate $\mathcal{A}^{(n+1)}$ by inserting $({I}^{(n+1)},\mathcal{N}^{(n+1)},\mathcal{A}^{(n)})$ into the right side of Eq.~(\ref{eq:Astar}).
\item Renormalize $\mathcal{N}^{(n+1)}$ and $\mathcal{A}^{(n+1)}$ as
$\mathcal{N}^{(n+1)} \to \mathcal{N}^{(n+1)}c$ and 
$\mathcal{A}^{(n+1)} \to \mathcal{A}^{(n+1)}/c$
with normalization factor $c = \sum_i\mathcal{A}_i^{(n+1)}$.
\item Repeat from step (ii) until the solution has sufficient
convergence, {i.e.}, the ratio of consecutive likelihoods~(\ref{eq:LH}) has $\Delta \chi^2 \simeq 2\ln(\mathcal{L}^{(n+1)}/\mathcal{L}^{(n)})\ll 1$.
\end{enumerate}
The local and celestial sky are binned following the {\tt HEALPix} parametrization of the unit sphere~\citep{Gorski:2004by}. 

The stability of the iterative reconstruction method for limited data can be improved by smoothing the local event distribution in each sidereal time bin before the start of the reconstruction. In the case of the Auger analysis above 8~EeV we chose a Gaussian symmetric beam with a full width half maximum of $4^\circ$. We have validated that this choice of smoothing scale does not affect the analysis of medium-scale anisotropies discussed in this analysis. Alternatively, one can reconstruct the anisotropy using the ansatz
\begin{equation}\label{eq:truncated}
{\delta I}(\alpha,\delta) = \sum^{\ell_{\rm max}}_{\ell\geq1}\sum_{m\neq 0} a_{\ell m} Y^{\ell m}(\pi/2-\delta,\alpha)\,,
\end{equation}
where $\ell_{\rm max}$ corresponds to the truncation scale that can be adjusted to a suitable value. With this ansatz, step {\it (iii)} in the iterative method is automatically implied. Under the condition of small anisotropies, $\delta I \ll 1$, we can solve the best-fit expansion coefficients $a^\star_{\ell m}$ with $m\neq0$ in terms of $\mathcal{N}^\star$ and $\mathcal{A}^\star$ by the following matrix equation:
\begin{multline}\label{eq:newI}
 \sum_{\ell'=1}^{\ell_{\rm max}}\sum_{m'\neq0}\left[\sum_{\mathfrak{a}}\omega_\mathfrak{a} (Y^{\ell m}_{\mathfrak{a}})^*Y^{\ell' m'}_{\mathfrak{a}}\right]a^\star_{\ell'm'}\\\simeq \sum_{\mathfrak{a}}\omega_\mathfrak{a}\left[1 - \frac{\sum_\tau{\mathcal{N}^\star_\tau\mathcal{A}^\star_{\tau\mathfrak{a}}}}{\sum_\sigma n_{\sigma \mathfrak{a}}}\right](Y^{\ell m}_{\mathfrak{a}})^*\,,
\end{multline}
where we again used the abbreviation $Y^{\ell m}_{\tau i} \equiv Y^{\ell m}({\bf R}(t_\tau){\bf n}'(\Omega_{i}))$ and $\omega_\mathfrak{a} \equiv {\sum_\tau n_{\tau \mathfrak{a}}}/N_{\rm tot}$. This expression (\ref{eq:newI}) determines the new anisotropy in step {\it (ii)} of the iterative methods. 

Expression (\ref{eq:newI}) has a familiar form. The term in parentheses on the right side of this equation is the anisotropy $\delta I^\star_\mathfrak{a}$ (cf.~Eq.~(\ref{eq:Istar})) in the limit $\delta I^\star \ll 1$. The right side is then simply the pseudo-coefficient $\widetilde{a}_{\ell m}$ of the anisotropy in the equatorial coordinate system with a weight function $\omega_\mathfrak{a}$. The matrix in parentheses on the left side of Eq.~(\ref{eq:newI}) relates this pseudo-coefficient $\widetilde{a}_{\ell m}$ to the true coefficients $a^\star_{\ell m}$. In the analyses of the cosmic microwave background, this expression is known as the coupling matrix $K_{\ell m\ell'm'}$ for the weight function $\omega_\mathfrak{a}$ (see, {\it e.g.}, ~\cite{Efstathiou:2003dj}). Note, that this coupling matrix can only be inverted to solve Eq.~(\ref{eq:newI}) for the $a^\star_{\ell m}$ if we truncate the spherical harmonic expansion at a finite $\ell_{\rm max}$. In the case of $\ell_{\rm max}=1$ we can rewrite Eq.~(\ref{eq:newI}) in terms of the real dipole components $d_x$ and $d_y$ as shown in Eq.~(\ref{eq:dxdystar}).

\section{Statistical Uncertainty}\label{AppendixC}

The statistical uncertainty of the large-scale anisotropy coefficients $a_{\ell m}$ from the ansatz (\ref{eq:truncated}) can be directly estimated from the likelihood function (\ref{eq:LH}). The inverse of the (normalized) covariance matrix is given as 
\begin{equation}
(V^{-1})_{{x_i}{x_j}} \equiv \left[x_ix_j\frac{\partial(-\ln \mathcal{L})}{\partial{x_i}\partial{x_j}}\right]_{{\bf x} = {\bf x}^\star}\,,
\end{equation}
where we introduced the parameter vector\footnote{Strictly speaking, the relative acceptance $\mathcal{A}_i$ is restricted to positive quantities and is subject to the normalization condition $\sum_i\mathcal{A}^\star_i = 1$. However, we will see in the following that the $a_{\ell m}$-$\mathcal{A}_i$ cross terms are negligible for the estimation of the statistical uncertainty of the large-scale anisotropy.} ${\bf x} = \lbrace a_{\ell m},\mathcal{A}_i,\mathcal{N}_\tau\rbrace$. 
We will first concentrate on matrix elements involving $a_{\ell m}$. The diagonal elements are given by
\begin{align}
(V^{-1})_{a_{\ell m}a_{\ell'm'}} &= \sum_{\tau i}n_{\tau i} I^{\ell m}_{\tau i} I^{\ell' m'}_{\tau i}\,,
\end{align}
where we use the abbreviation $I^{\ell m}_{\tau i} \equiv a^\star_{\ell m}Y^{\ell m}({\bf R}(t_\tau){\bf n}'(\Omega_{i}))$. The off-diagonal elements are of the form
\begin{align}\label{eq:crosstermN}
(V^{-1})_{a_{\ell m}\mathcal{N}_\tau} &\simeq\mathcal{N}^\star_\tau\sum_i  \mathcal{A}^\star_iI^{\ell m}_{\tau i}\,,\\\label{eq:crosstermA}
(V^{-1})_{a_{\ell m}\mathcal{A}_i} &\simeq \mathcal{A}^\star_i\sum_\tau\mathcal{N}^\star_\tau I^{\ell m}_{\tau i}\,.
\end{align} 
The $a_{\ell m}$-$\mathcal{N}_\tau$ element (\ref{eq:crosstermN}) corresponds to an integral of the anisotropy $I^{\ell m}$ observed at a local sidereal time step $\tau$ over the local field of view weighted by the detector acceptance $\mathcal{A}^\star$. Due to the partial sky coverage, this matrix element is expected to have a sizable contribution. On the other hand, the $a_{\ell m}$-$\mathcal{A}_i$ element (\ref{eq:crosstermA}) corresponds to an integral of the anisotropy $I^{\ell m}$ observed at a fixed local position $i$ over local sidereal time and weighted by the background rate $\mathcal{N}^\star$. For $m\neq0$ spherical harmonics and near-uniform background rates, this matrix element is not expected to have a large contribution. As a consequence, we will evaluate the covariance matrix for the reduced parameter set ${\bf x} = \lbrace a_{\ell m},\mathcal{N}_\tau\rbrace$. The remaining matrix element is simply
\begin{equation}
(V^{-1})_{\mathcal{N}_\tau \mathcal{N}_\sigma} = \delta_{\tau\sigma}\sum_in_{\tau i}\,.
\end{equation}
After integration of the likelihood over the uncertainty of the background rate $\mathcal{N}_\tau$ we can write the marginalized covariance matrix $\overline{V}$ of the $\lbrace{a_{\ell m}}\rbrace$ parameters as
\begin{multline}
(\overline{V}^{-1})_{a_{\ell m}a_{\ell'm'}} \simeq \sum_{\tau i}n_{\tau i} I^{\ell m}_{\tau i} I^{\ell' m'}_{\tau i}\\-\sum_\tau \frac{(\mathcal{N}^\star_\tau)^2}{\sum_in_{\tau i}}\left(\sum_i\mathcal{A}^\star_i I^{\ell m}_{\tau i}\right)\left(\sum_j\mathcal{A}^\star_j I^{\ell' m'}_{\tau j}\right)\,.
\end{multline}
In the case of the dipole anisotropy reconstruction ($\ell_{\rm max }=1$), this matrix is approximately diagonal and we can write the uncertainty of the dipole components $d_{x/y}$ as in Eq.~(\ref{eq:sigma}).

\begin{figure}[t]\centering
\includegraphics[width=\linewidth]{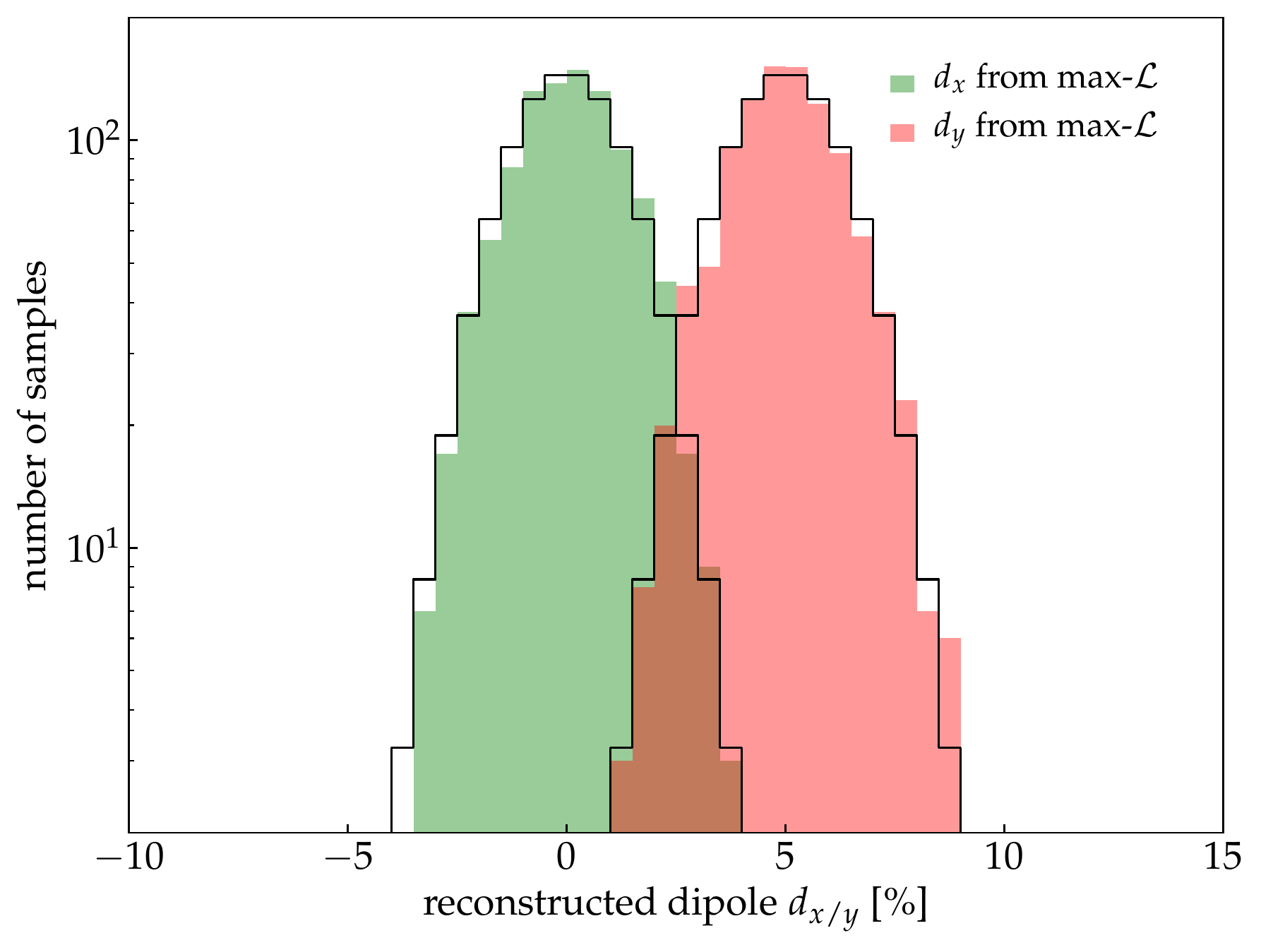}
\caption[]{Reconstructed dipole anisotropy for 1000 samples of mock data. The filled histograms show the distribution of reconstructed results for $d_x$ (green) and $d_y$ (red) around the true input values $d^{\rm true}_x=0$ and $d^{\rm true}_y=5$\%. The predicted variance of the distribution (\ref{eq:sigma}) is $\sigma_x\simeq\sigma_y\simeq1.35$\% and is indicated by the open histograms.}\label{fig:validation}
\end{figure}

In order to validate the dipole reconstruction method and corresponding parameter estimation we simulate 1000 mock data samples based on a pure dipole anisotropy with $d^{\rm true}_x=0$ and $d^{\rm true}_y=5$\%. For each simulation we use the same best-fit background rate $\mathcal{N}^\star$ and relative acceptance $\mathcal{A}^\star$ from the full anisotropy construction of the Pierre Auger data. We then use the iterative method outlined in Appendix~\ref{AppendixB} to reconstruct the dipole components $d_x$ and $d_y$. Each simulation arrives at a statistical uncertainty of the components with $\sigma_x\simeq\sigma_y\simeq1.35$\%. Figure~\ref{fig:validation} shows the distribution of reconstructed dipole coefficients in comparison to Gaussian distributions centered on $d^{\rm true}_{x/y}$ with a width of $1.35$\%. The mock data follow the predicted distributions well.
\newpage
\bibliographystyle{apj}
\bibliography{AugerReco}

\end{document}